\documentclass[twocolumn, pre, showpacs, english, preprintnumbers, amsmath, amssymb, superscriptaddress, aps,longbibliography]{revtex4-2}
\usepackage[subpreambles=true]{standalone}

\usepackage[utf8]{inputenc}
\usepackage{graphicx}
\usepackage{grffile}
\usepackage[usenames,dvipsnames]{xcolor}
\usepackage{amsmath}
\usepackage[normalem]{ulem}
\usepackage[resetlabels, labeled]{multibib}
\usepackage{appendix}
\newcites{S}{References Supplementary Materials}
\definecolor{orange}{rgb}{1,0.5,0}
\definecolor{goodgreen}{rgb}{0.1,0.5,0}
\definecolor{goodred}{rgb}{0.7,0,0}

\usepackage{tikz}
\usepackage{tocvsec2}
\usepackage{scrextend}
\usepackage{lineno}
\modulolinenumbers[5]
\makeatletter

\usepackage[colorlinks,urlcolor=goodgreen,citecolor=blue,linkcolor=goodred]{hyperref}

\usepackage{float}
\graphicspath{ {images/} }

\let\oldepsilon\epsilon \let\epsilon\varepsilon \let\varepsilon\oldepsilon

\makeatother

\usepackage{babel}

\begin{document}

\title{Coexistence of superconductivity and spin-splitting fields in superconductor/ferromagnetic insulator bilayers of arbitrary thickness}

\newcommand{\orcid}[1]{\href{https://orcid.org/#1}{\includegraphics[width=8pt]{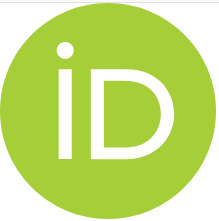}}}

\author{Alberto Hijano\orcid{0000-0002-3018-4395}}
\email{ahijano001@ikasle.ehu.eus}
\affiliation{Centro de F\'isica de Materiales (CFM-MPC) Centro Mixto CSIC-UPV/EHU, E-20018 Donostia-San Sebasti\'an,  Spain}

\author{Stefan Ili\'{c}}
\email{stefan.ilic@csic.es}
\affiliation{Centro de F\'isica de Materiales (CFM-MPC) Centro Mixto CSIC-UPV/EHU, E-20018 Donostia-San Sebasti\'an,  Spain}

\author{Mikel Rouco\orcid{0000-0003-2175-9238}}
\affiliation{Centro de F\'isica de Materiales (CFM-MPC) Centro Mixto CSIC-UPV/EHU, E-20018 Donostia-San Sebasti\'an,  Spain}

\author{Carmen Gonz\'{a}lez-Orellana}
\affiliation{Centro de F\'isica de Materiales (CFM-MPC) Centro Mixto CSIC-UPV/EHU, E-20018 Donostia-San Sebasti\'an,  Spain}

\author{Maxim Ilyn}
\email{maxim.ilin@ehu.eus}
\affiliation{Centro de F\'isica de Materiales (CFM-MPC) Centro Mixto CSIC-UPV/EHU, E-20018 Donostia-San Sebasti\'an,  Spain}

\author{Celia Rogero}
\affiliation{Centro de F\'isica de Materiales (CFM-MPC) Centro Mixto CSIC-UPV/EHU, E-20018 Donostia-San Sebasti\'an,  Spain}
\affiliation{Donostia International Physics Center (DIPC), 20018 Donostia--San Sebasti\'an, Spain}

\author{P. Virtanen}
\affiliation{Department of Physics and Nanoscience Center, University of Jyväskylä, P.O. Box 35 (YFL), FI-40014 University of Jyväskylä, Finland}

\author{T. T. Heikkilä}
\affiliation{Department of Physics and Nanoscience Center, University of Jyväskylä, P.O. Box 35 (YFL), FI-40014 University of Jyväskylä, Finland}

\author{S. Khorshidian\orcid{0000-0002-2374-0728}}
\affiliation{Department of Physics, College of Sciences, Yasouj University, Yasouj, 75914-353, Iran}
\affiliation{NEST, Istituto Nanoscienze-CNR and Scuola Normale Superiore, I-56127 Pisa, Italy}

\author{M. Spies\orcid{0000-0002-3570-3422}}
\affiliation{NEST, Istituto Nanoscienze-CNR and Scuola Normale Superiore, I-56127 Pisa, Italy}

\author{N. Ligato}
\affiliation{NEST, Istituto Nanoscienze-CNR and Scuola Normale Superiore, I-56127 Pisa, Italy}

\author{F. Giazotto\orcid{0000-0002-1571-137X}}
\affiliation{NEST, Istituto Nanoscienze-CNR and Scuola Normale Superiore, I-56127 Pisa, Italy}

\author{E. Strambini\orcid{0000-0003-1135-2004}}
\email{elia.strambini@sns.it}
\affiliation{NEST, Istituto Nanoscienze-CNR and Scuola Normale Superiore, I-56127 Pisa, Italy}

\author{F. Sebasti\'{a}n Bergeret\orcid{0000-0001-6007-4878}}
\email{fs.bergeret@csic.es}
\affiliation{Centro de F\'isica de Materiales (CFM-MPC) Centro Mixto CSIC-UPV/EHU, E-20018 Donostia-San Sebasti\'an,  Spain}
\affiliation{Donostia International Physics Center (DIPC), 20018 Donostia--San Sebasti\'an, Spain}
\affiliation{Institute of Solid State Theory, University of Münster, D-48149 Münster, Germany}

\begin{abstract}
Ferromagnetic insulators (FI) can induce a strong exchange field in an adjacent superconductor (S) via the magnetic proximity effect. This manifests as spin-splitting of the BCS density of states of the superconductor, an important ingredient for numerous superconducting spintronics applications and the realization of Majorana fermions.
A crucial parameter that determines the magnitude of the induced spin-splitting in FI/S bilayers is the thickness of the S layer $d$: In very thin samples, the superconductivity is suppressed by the strong magnetism. By contrast, in very thick samples, the spin splitting is absent at distances away from the interface. In this work, we calculate the density of states and critical exchange field of FI/S bilayers of arbitrary thickness. From here, we determine the range of parameters of interest for applications, where the exchange field and superconductivity coexist. We show that for  $d>3.0\xi_s$, the paramagnetic phase transition is always of the second order, in contrast to the first-order transition in thinner samples at low temperatures. Here $\xi_s$ is the superconducting coherence length. Finally, we compare our theory with the tunneling spectroscopy measurements in several  EuS/Al/AlO$_x$/Al samples. If the  Al film in contact with the EuS  is thinner than a certain critical value, we do not observe superconductivity, whereas, in thicker samples, we find evidence of a first-order phase transition induced by an external field. The complete transition is preceded by a regime in which normal and superconducting regions coexist. We attribute this mixed phase to inhomogeneities of the Al film thickness and the presence of superparamagnetic grains at the EuS/Al interface with different switching fields. The step-like evolution of the tunnel-barrier magnetoresistance supports this assumption.
Our results demonstrate on the one hand, the important role of the S layer thickness, which is particularly relevant for the fabrication of high-quality samples suitable for applications. On the other hand, the agreement between theory and experiment demonstrates the accuracy of our theory, which, originally developed for homogeneous situations, is generalized to highly inhomogeneous systems. 
\end{abstract}

\maketitle

\section{Introduction}

It was shown a long time ago~\cite{moodera1988electron}, and confirmed in several later experiments~\cite{Hao:1990,Hao:1991,Moodera_2007,Miao:2009,Xiong:2011,li2013observation,Strambini:2017,DeSimoni:2018,rouco2019charge}, that a thin superconducting film (S), adjacent to a ferromagnetic insulator (FI),  may exhibit a spin-split density of states even at zero field. The splitting is due to the interfacial exchange interaction  between the localized magnetic moments and the Al film's conduction band electrons. Even though back in the late 80s and 90s, this  effect had attracted attention mainly from a fundamental research perspective~\cite{meservey1994spin}, only recently superconductors with a spin-split density of states (DoS)  are proposed for diverse applications, such as topological qubits using Majorana wires~\cite{Oreg:2010,lutchyn2010majorana}, spin valves \cite{miao2014spin,DeSimoni:2018}, thermometry~\cite{Giazotto:2006,Giazotto:2015}, magnetometers~\cite{Alidoust:2013,strambini2015mesoscopic}, caloritronic devices~\cite{giazotto2020very,giazotto2015very,giazotto2013phase}, thermoelectricity \cite{machon2013nonlocal,ozaeta2014predicted}, and radiation detectors~\cite{Heikkila:2018}. 

An ideal material combination for observing spin-split superconductivity at zero field is EuS/Al.  This has been confirmed in numerous spectral measurements on EuS/Al samples, mainly grown by Moodera's group at the MIT~\cite{Hao:1990,Moodera_2007,Strambini:2017,DeSimoni:2018}. It is understood that the splitting size at the interface is proportional to the interfacial exchange field, which in turn is proportional to the averaged magnetic moment of the EuS~\cite{Zhang:2019,Strambini:2017}. Thus, one needs high-quality S/FI  interfaces for a sizable exchange field, avoiding a non-magnetic interlayer between the two materials. It is also known that the effective splitting field decays away from the interface over the superconducting coherence length~\cite{Tokuyasu:1988}. Thus, for applications that require an almost homogeneous splitting, the S layers have to be thin enough. On the other hand, the induced exchange cannot be too strong because it would destroy the superconductivity~\cite{Chandrasekhar:1962,Clogston:1962}. The difficulty then lies in manufacturing superconducting films thin enough to have a sufficiently large splitting field, but at the same time, the field has to be weak enough in order not to suppress the superconducting state. 

Indeed, FI/S systems with a  S layer thinner than the superconducting coherence length behave as homogeneous superconductors in a Zeeman field. In this case, the well-established theory of a paramagnetic phase transition to the normal state applies \cite{Sarma:1963,Maki-Tsuneto:1964}. However, if the S layer's thickness is comparable to the superconducting coherence length, the spin-splitting field is non-homogeneous, and hence the theory needs to be revised. The new theory has to connect the thin layer limit, in which the phase transition takes place, and the thick S layer limit,  in which one expects no transition to the normal state for any value of the interfacial exchange field.  Clearly, in this latter case, the splitting is negligible at the boundary opposite to the FI/S interface, and hence such a system is less suitable for applications requiring spin-splittings. It is crucial for experiments to find the optimal values of the interfacial exchange field and the S-layer's thickness to simultaneously observe a well-defined superconducting gapped state and a sharp spin splitting of the quasiparticle peaks at zero field.
Even though several works have studied the effect of a homogeneous spin-splitting field on the superconducting properties of the S layer in FI/S structures,
there is no study, to the best of our knowledge, on the role of the S thickness on the spectral properties of FI/S junctions.~\footnote{ In Refs. \cite{Mironov:2012,Mironov:2018} a possible  phase transition to the FFLO state has been studied  in all-metallic ferromagnet-superconductor and superconductor-ferromagnet-normal metal structures with different thickness. According to these works, the FFLO may appear when the conductivity of the non-superconducting region is much larger than the  conductivity of S in the normal state. In our case we consider a ferromagnetic insulator and hence we are in the opposite limit.  Therefore, we may ignore the FFLO state \cite{aslamazov1969influence,Virtanen:2020} }

This work addresses this problem and presents an exhaustive theoretical analysis of the spectral properties and phase transition of diffusive superconducting films of arbitrary thickness adjacent to a FI layer.
The combination of the DoS and the phase diagram gives a complete picture of the system that can help to identify the range of parameters where superconductivity and a well-defined spin splitting coexist, which is the desired situation for applications. Moreover, we infer the nature of the phase transition at different temperatures from the non-monotonic behavior of the critical exchange field and find a temperature-dependent critical thickness above which there is no phase transition, regardless of the value of the exchange field.
Our work also includes the fabrication and transport measurements of EuS/Al/AlO$_\mathrm{x}$/Al junctions.  The Al film next to the EuS layer exhibits a spin-split DoS. By applying an external magnetic field the spin-splitting changes.  At 30 mK we observe a clear signature of a first-order phase transition when the splitting field equals to $\Delta_0^*/\sqrt{2}$, where $\Delta_0^*$ is the superconducting gap at zero exchange field in the presence of magnetic impurities. We also show the magnetic field dependence of the resistance at very low temperatures, which suggests the presence of a mixed phase with superconducting and normal parts at certain field ranges.  This mixed phase is confirmed by the good agreement between the measured differential conductance and the results of our theoretical model. 

This paper is organized as follows:  In Sec.~\ref{model} we present the basic equations describing diffusive superconductors and the boundary conditions at the FI/S interface. In Sec.~\ref{DOS section} we obtain the density of states (DoS) for different values of the thickness of the S layer. In Sec.~\ref{critical field section} we calculate the critical exchange field of the system and discuss the nature of the paramagnetic phase transition. In Sec.~\ref{Comparison to experiments}, we present details of the fabrication of EuS/Al/AlO$_\mathrm{x}$/Al junctions and the tunneling spectroscopy measurements, which we compare to our theoretical results. We summarize the results in Sec.~\ref{conclusion}.

\section{Model and formalism}\label{model}

\begin{figure}[!t]
    \centering
    \includegraphics[width=0.99\columnwidth]{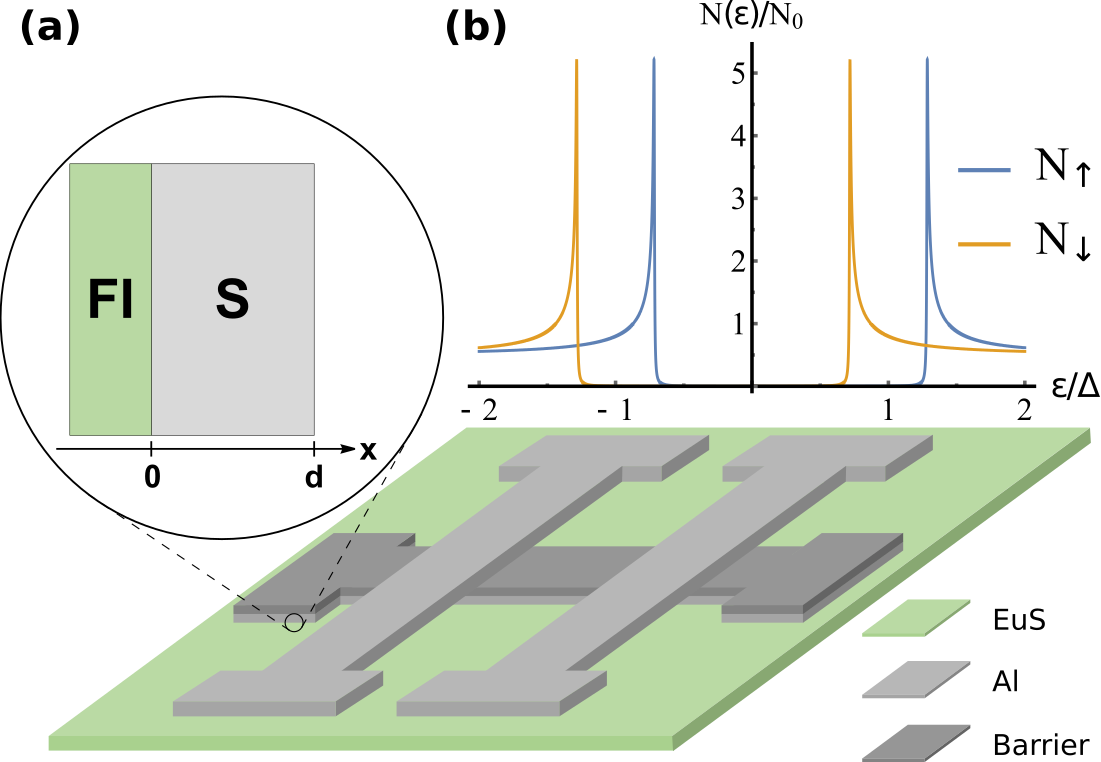}
    \caption{(a) Experimental setup and schematic view of the FI/S bilayer. The S layer in contact with the FI has a thickness $d$. See Sec.~\ref{Comparison to experiments} for additional details on the experiment. (b) DoS of a homogeneous spin-split superconductor.}\label{FI/S scheme}
\end{figure}

In this section, we introduce the basic equations determining  the spectral properties of a FI/S bilayer [see the inset of Fig.~\ref{FI/S scheme}(a)] for arbitrary values of the exchange field and S layer thickness $d$. We assume that the system is homogeneous in the directions parallel to the interface.

\begin{figure*}
\centering
  \includegraphics[width=0.99\textwidth]{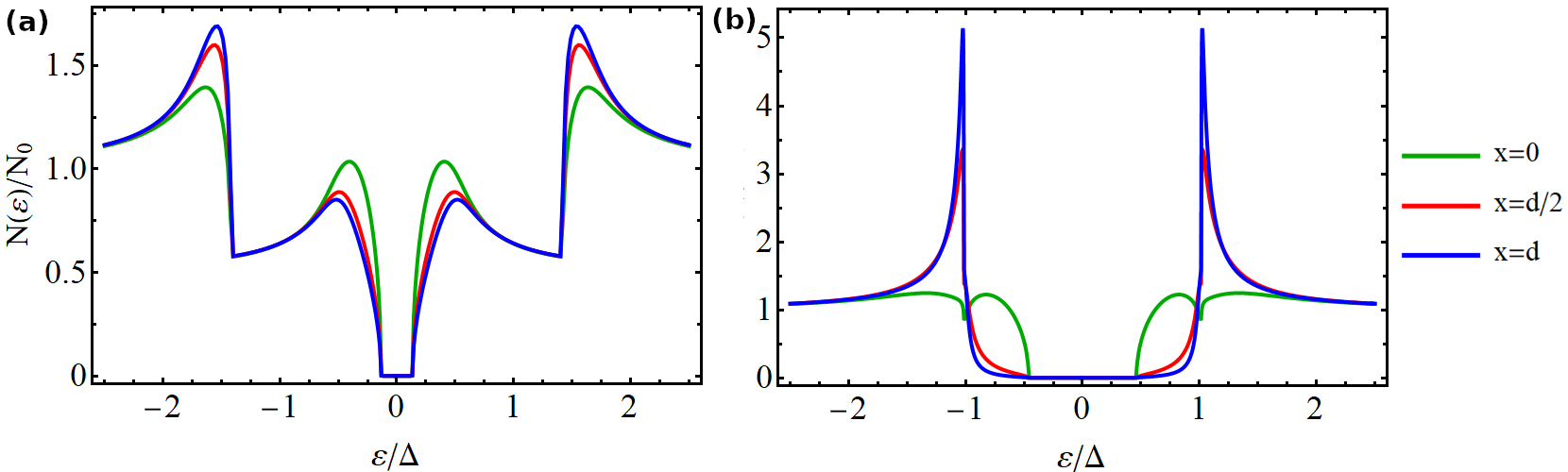}
  \caption{DoS for (a) a superconductor of intermediate thickness ($d=0.5\xi_0$) and (b) thick superconductor ($d=3\xi_0$) at different distances from the FI/S interface: $x=0$ (green), $x=d/2$ (red) and $x=d$ (blue). Note that the energy is normalized by the order parameter. The value of the exchange field is $ha/\xi_0=0.3\Delta$, and we assume no magnetic impurities ( $\tau_{\rm sf}^{-1}=0$). }\label{nDOS_total}
\end{figure*}

We describe the system using the Green’s function (GF) technique. In the case under consideration, the spatial scales over which the superconducting properties vary are much larger than the Fermi wavelength. Moreover, all relevant energies involved are smaller than the Fermi energy. In this case, one can use the quasiclassical approximation, which simplifies the problem considerably in hybrid systems~\cite{Eilenberger,Larkin-Ovchinnikov:1968,Usadel,volkov1993proximity,lambert1998phase,belzig1999quasiclassical}.
Because we are dealing with superconductivity and spin-dependent fields, the quasiclassical Green's function $\check{g}$ is a $4 \times 4$ matrix in Nambu-spin space. The density of states is related to the retarded GF $\check{g}^R$ as
\begin{equation}\label{DOSformula}
    N(\boldsymbol{r},\epsilon)=\frac{N_0}{4}\mathrm{Re}\, \mathrm{Tr}\{\tau_3 \check{g}^R(\boldsymbol{r},\epsilon)\}\; ,
\end{equation}
where $N_0$ is the density of states at the Fermi level in the normal state and $\mathrm{Tr}$ denotes the trace over Nambu and spin spaces.

The GF in Eq.~\eqref{DOSformula} has to be determined from the quasiclassical equations, which in the dirty limit  reduce to a diffusive-like equation, known as the Usadel equation~\cite{Usadel} \footnote{ Here and throughout the paper we set $\hbar=1$.}, 
\begin{equation}\label{usadel0}
    D \partial_x(\check{g}\partial_x\check{g})+[i\epsilon\tau_3+\Delta\tau_2-i\boldsymbol{h}\cdot\boldsymbol{\sigma}\tau_3-\check{\Sigma},\check{g}]=0\; ,
\end{equation}
and the normalization condition, $\check{g}^2=\check{1}$. 
Here, $D$ is the diffusion constant, $\epsilon$ is the energy, $\Delta$ is the superconducting order parameter, and $\boldsymbol{h}$ is the exchange field.  The latter is only finite at the FI/S interface, and we approximate it as  $\boldsymbol{h}=ha\delta(x)\hat{\boldsymbol{z}}$, where $h$ is the exchange field, and $a$ is the thickness of an effective layer over which the exchange interaction is finite  \cite{Zhang:2019}. The matrices  $\sigma_i$ and $\tau_i$ ($i=1,2,3$) in Eq.~\eqref{usadel0}  are the Pauli matrices in spin and Nambu space, respectively.
$\check{\Sigma}$ is the self-energy term, which describes different scattering  processes, such as magnetic and spin-orbit impurities~\cite{meservey1994spin,bergeret2018colloquium,HEIKKILA2019100540}. Most of the experiments on spin-split superconductors, including those in the present work, are made using Al layers, for which the spin-orbit coupling can be neglected. Therefore we only  consider the spin-flip relaxation processes, described by  the self-energy term: 
\begin{equation}\label{sf relaxation}
    \check{\Sigma}_{\mathrm{sf}}=\frac{\sigma_i\tau_3\check{g}\tau_3\sigma_i}{8\tau_{\mathrm{sf}}}a\delta(x)\; ,
\end{equation}
where $\tau_{\mathrm{sf}}$ is the spin-flip relaxation time, and sum over repeated indices is implied. Even though the Al films often have a tiny concentration of homogeneously distributed magnetic impurities, the main source of magnetic disorder and a sizable spin-flip relaxation is the FI/S interface~\cite{meservey1994spin}. This assumption is supported by contrasting our model with the experimental data presented  in Sec.~\ref{Comparison to experiments}. As with the exchange field, we then assume that the self-energy, Eq.~\eqref{sf relaxation}, is only finite at the interface. Thus, the Usadel equation~\eqref{usadel0} in the superconducting layer does not contain spin-dependent terms and becomes
\begin{equation}\label{usadel1}
    D \partial_x(\check{g}\partial_x\check{g})+[i\epsilon\tau_3+\Delta\tau_2,\check{g}]=0\; .
\end{equation}
Both the exchange and spin-relaxation terms enter the boundary conditions at the FI/S interface ($x=0$), which can be obtained by integrating Eq. (\ref{usadel0}) in a small region around the interface: \begin{equation}\label{boundary condition2}
    \left. \check{g}\partial_x\check{g}\right|_{x=0}=\frac{1}{D}\left. \left[iha\sigma_3\tau_3+\frac{\sigma_i\tau_3\check{g}\tau_3\sigma_i}{8\tau_{\mathrm{sf}}}a,\check{g}\right]\right|_{x=0}\; .
\end{equation}
The second term in the commutator stems from the spin-flip processes and  mixes GF components with opposite spins.

%
%

The spectral current  vanishes at the boundary with vacuum or an insulator. This implies the boundary condition
\begin{equation}\label{boundary condition1}
\begin{split}
\left. \check{g}\partial_x\check{g}\right|_{x=d}=0\; .
\end{split}
\end{equation}
The quasiclassical GF is then determined from  Eqs. (\ref{usadel1}-\ref{boundary condition1}) and the normalization condition. This set of equations is complemented by the self-consistency relation for $\Delta$. 
In Secs.~\ref{DOS section} and \ref{critical field section}, we solve this boundary problem numerically and compute the DoS and the critical temperature and critical exchange field for FI/S bilayers,  for arbitrary values of the interfacial exchange field and S layer thickness.

\section{Density of states of a FI/S bilayer}\label{DOS section}
In this section, we study the spatial dependence of the DoS for different thicknesses of the superconducting layer. We solve the boundary problem (\ref{usadel1}-\ref{boundary condition1}) and evaluate the DoS using Eq.~\eqref{DOSformula}.

\begin{figure*}
\centering
  \includegraphics[width=0.99\textwidth]{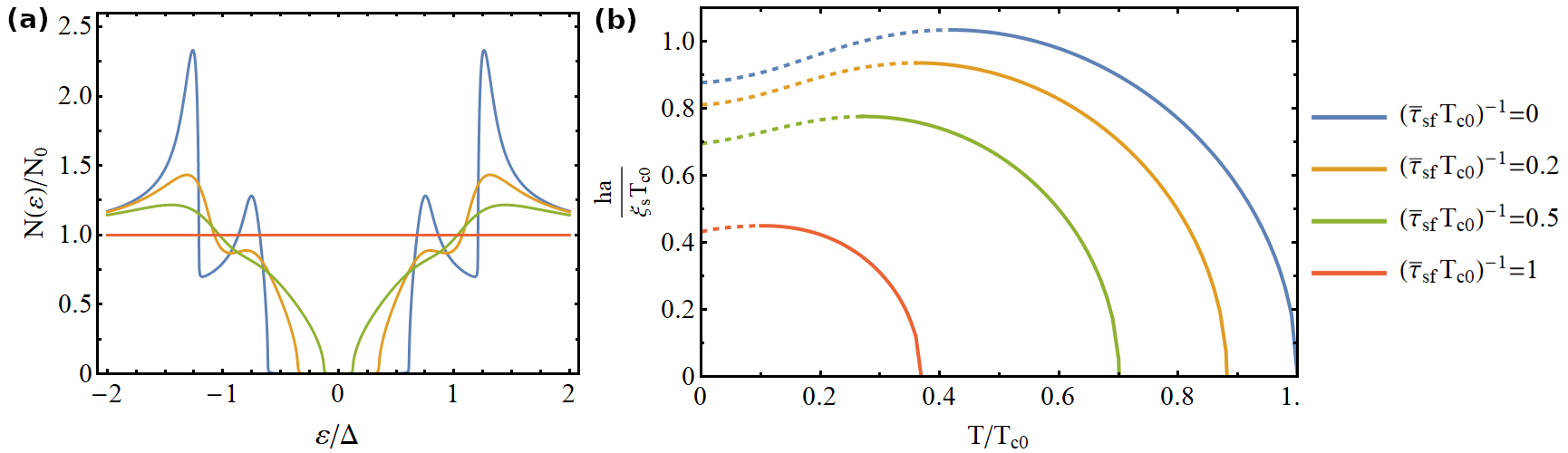}
  \caption{(a) DoS at $x=d$ for different values of the spin-flip relaxation rate. The samples have a thickness $d=\xi_s$, and the value of the exchange field is $ha/\xi_s=0.5T_{c0}$. (b) Phase boundary for the second order phase transition. The solid part of the lines represent the critical exchange field for the second order transition, while the dashed lines correspond to the temperature range where the first order phase transition occurs.}\label{critical field magnetic impurities}
\end{figure*}

We start analyzing the thin-film limit. 
The characteristic length scale of the Usadel equation is the superconducting coherence length, which at low temperatures is approximately given by $\xi_0=\sqrt{D/\Delta}$. In the thin-layer limit the thickness of the superconductor $d$ is much smaller than $\xi_0$, so we may assume that $\check{g}$ is constant. The effective value for the exchange field, $\bar h$  and the spin-flip rate, $1/\bar \tau_{\mathrm{sf}}$ can then be obtained by integrating Eq.~\eqref{usadel1} over the Al thickness and the boundary conditions Eqs.~(\ref{boundary condition2}-\ref{boundary condition1})
\begin{eqnarray}
\bar h&=&ha/d,\label{barh}\\
\bar\tau_{\rm sf}^{-1}&=&\tau_{\rm sf}^{-1} a/d.\label{bartau}
\end{eqnarray}
In the absence of spin-flip scattering, $\bar\tau_{\rm sf}^{-1}=0$, the spin-resolved DoS for spin-up ($\uparrow$) and spin-down ($\downarrow$) quasiparticles has the form  as for a homogeneous superconductor in a Zeeman field:
\begin{equation}\label{HomDos}
N_{\uparrow,\downarrow}(\epsilon,\bar{h})=\frac{N_0}{2}\mathrm{Re}\frac{|\epsilon\mp \bar{h}|}{\sqrt{(\epsilon\mp \bar{h})^2-\Delta^2}}.
\end{equation}

In Fig.~\ref{FI/S scheme}(b) we show the corresponding DoS of a homogeneous spin-split superconductor given by this equation. The homogeneous exchange field $\bar{h}$ induces a spin splitting of the density of states, such that the states for each spin direction are raised or lowered in energy.

The  DoS of  superconductors of intermediate ($d=0.5\xi_0$) and large thicknesses ($d=3\xi_0$) are calculated numerically, and shown in Fig.~\ref{nDOS_total}. Unlike the homogeneous thin layer limit, the DoS in this case varies in space. For intermediate S layers, the spin-splitting remains almost constant along the sample, but the spin-splitting vanishes away from the FI/S interface in thick samples. This is as expected, since in the thick sample limit we should recover the BCS density of states without any spin-splitting at distances far away from the interface. Another notable effect is that as we move away from the interface, the height of the inner peaks is reduced, while the outer peaks increase in size.

Comparing panels (a) and (b) in Fig.~\ref{nDOS_total} we see that for a given value of the interfacial exchange field $h$, the spin-splitting decreases by increasing thickness of the S layer. The value of the spin-splitting in the very thin superconductor limit is given by $2 \bar{h}$, where $\bar{h}$ is the average exchange field defined in Eq.~\eqref{barh}. Extracting the spin-splitting from the panels, we show that the splitting at $x=0$ is well approximated by this expression even for thick samples.

In Fig.~\ref{critical field magnetic impurities}(a) we show the DoS of a S layer of intermediate thickness for different values of the effective spin-flip rate [see Eq.~\eqref{bartau}]. As the spin-flip rate is increased, the coherence peaks are smeared and the superconducting gap is suppressed. Above a threshold value of the spin-flip rate, the superconductor is driven to the normal state.

Increasing the value of the exchange field beyond some threshold value induces a phase transition from the superconducting state to the normal state. We analyze this transition in the next section.

\section{Critical temperature of a FI/S bilayer}\label{critical field section}

In this section we determine the critical temperature $T_c$ of the FI/S bilayers. In addition to the DoS, this is another experimentally accessible quantity, which is determined by measuring the resistance drop as a function of temperature.

\begin{figure*}
\centering
  \includegraphics[width=1.0\textwidth]{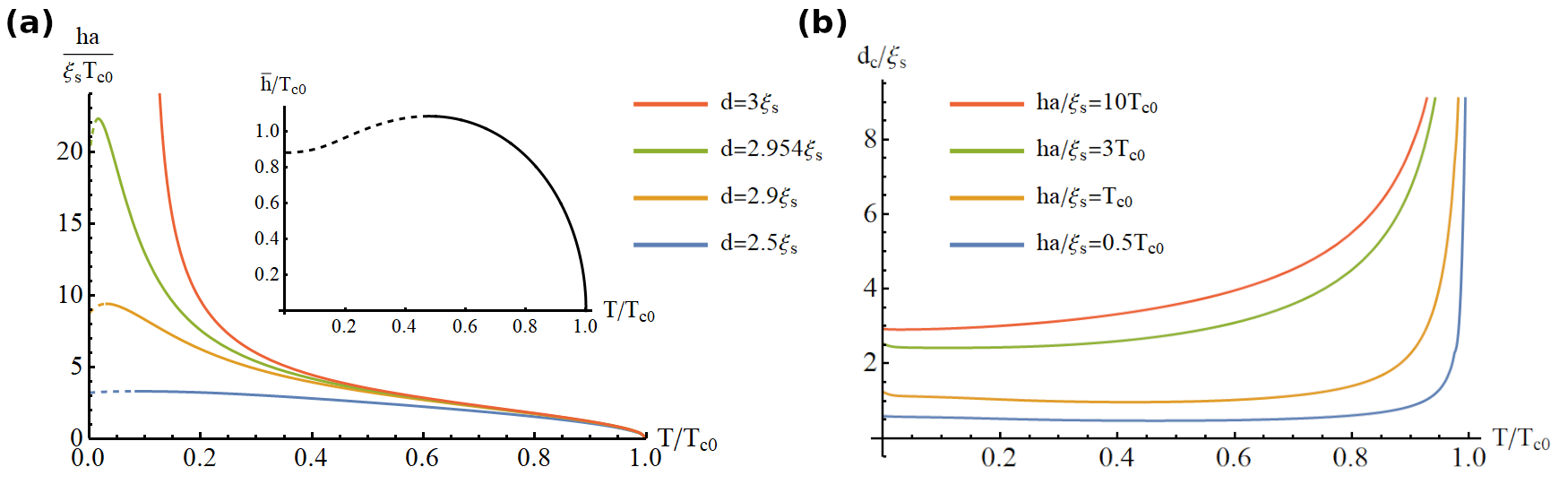}
  \caption{(a) Critical exchange field for different thicknesses of the superconductor and (b) temperature dependence of the critical thickness for different exchange fields in the magnetic impurity-free limit ($\tau_{\mathrm{sf}}^{-1}=0$). The inset in panel (a) corresponds to the critical exchange field in the thin limit (homogeneous case). The dashed part corresponds to the temperature region for which the transition is of first order.}\label{multimode}
\end{figure*}

Calculation of $T_c$ as a function of other system's parameters allows us to construct the phase-diagram for a FI/S bilayer. For this sake we employ the Matsubara Green's functions, which are obtained from Eqs.~(\ref{usadel1}-\ref{boundary condition1}) after the substitution $\epsilon \rightarrow i\omega_n$. Here, $\omega_n=2\pi T(n+1/2)$, $n \in \mathbb{Z}$, are the Matsubara frequencies~\cite{Abrikosov-Gorkov-Dzyaloshinski}.

First,  we assume a second order phase transition and look for the  solution of the Usadel equation at temperatures close to $T_c$.  In this case one can linearize the problem by treating  $\Delta$ perturbatively. Near $T_c$ the GF can be approximated by $\check g=\mathrm{sgn}(\omega_n)\tau_3 +i\hat f \tau_2$, where the anomalous function $\hat{f}$ satisfies the linearized Usadel equation: 
\begin{equation}\label{Usadel_linear1}
    \xi_s^2 \pi T_{c0}\partial_{xx}^2 \hat{f}-|\omega_n|\hat{f}+i\Delta=0\; ,
\end{equation}
with $\xi_s^2=D/(2 \pi T_{c0})$, where $T_{c0}$ is the critical temperature in the absence of the magnetic field and $\xi_s$ is the coherence length close to the critical temperature.

As discussed in Sec. \ref{model}, the exchange field and spin-flip scattering rate enter the boundary condition at the FI/S interface which after linearization reads [see Eq.~\eqref{boundary condition2}]
\begin{subequations}\label{Usadel_linear_boundary}
\begin{equation}
    \partial_x \left.\hat{f}\right|_{x=0}=i\:\mathrm{sgn}(w_n)\kappa_h \sigma_3 \left.\hat{f}+\kappa_{sf} \left(3\hat{f}+\sigma_i\hat{f}\sigma_i\right)\right|_{x=0}\; ,
\end{equation}
\begin{equation}
    \partial_x \left.\hat{f}\right|_{x=d}=0\; ,
\end{equation}
\end{subequations}
where $\kappa_h=ah/(\pi T_{c0}\xi_s^2)$ and $\kappa_{sf}=a/(8\tau_\mathrm{sf}\pi T_{c0}\xi_s^2)$.
The order parameter $\Delta$ is then determined self-consistently from the  gap equation~\cite{Kopnin:2001}:  
\begin{equation}\label{self-consistency0}
    \Delta\ln{\frac{T_{c0}}{T}}=\pi T\sum_{\omega_n}\left(\frac{\Delta}{|\omega_n|}+\frac{i}{2}\mathrm{Tr} \hat{f}\right)\; .
\end{equation}
From this equation we can determine the critical temperature and field. 
In order to solve this problem we use the ansatz for $\hat{f}$ prescribed by the so-called multimode method developed by Fominov \textit{et al.}~\cite{Fominov:2002} (see Appendix \ref{Multimode method ansatz} for details).

The nature of the paramagnetic phase transition of a FI/S bilayer depends on the temperature, but also on the degree of disorder of the metal. Namely, Tokuyasu \emph{et al.}~\cite{Tokuyasu:1988} have shown that in the clean limit, this phase transition is always of the second order. However, a recent work established that even a small amount of impurity scattering restores the possibility of a first-order phase transition~\cite{Virtanen:2020}. We assume here the diffusive limit, which is well justified for Al films due to their intrinsic disorder.

In the thin layer limit, $d\ll \xi_s$, FI/S bilayers are homogeneous, and evaluating the gap equation \eqref{self-consistency0} yields the standard expression for the paramagnetically limited second-order phase transition in the absence of magnetic impurities~\cite{Kopnin:2001}
\begin{equation}
\ln \frac{T_{c0}}{T}=\mathrm{Re}\bigg[\psi \bigg(\frac{1}{2}+\frac{i\bar{h}}{2\pi T}\bigg)-\psi \bigg(\frac{1}{2}\bigg)\bigg]\; .
\label{critfield}
\end{equation}
Here, $\psi$ is the digamma function. Importantly, Eq.~\eqref{critfield} only holds when the temperature is higher than $T^*=0.56 T_{c0}$. At lower temperatures, the phase transition is of the first order, which can be readily proved by analyzing the free energy of the superconductor~\cite{Kopnin:2001}. At $T=0$, the critical field is given by the Chandrasekhar-Clogston limit $\bar{h}=\Delta_0/\sqrt{2}$~\cite{Chandrasekhar:1962,Clogston:1962}. The nature of the phase transition can, however, be inferred even without knowing the free energy, from the shape of the $\bar{h}(T)$ critical line. Namely, below $T^*$ the critical line exhibits non-monotonic behavior, and the critical temperature is a double-valued function of the field [see Fig.~\ref{multimode}(a)], but the smaller solution is physically unstable. Therefore, the onset of the first-order phase transition coincides with the non-monotonic features in the $h(T)$ diagram. We assume that the multivaluedness of the solution surface $\Delta(h,T)$ is correlated with the nonmonotonicity of $h(T)$ to identify the nature of the phase transition in thick FI/S bilayers.


We first consider the effect of spin-flip relaxation. In Fig.~\ref{critical field magnetic impurities} we show the DoS and $h(T)$ diagram of a S layer of intermediate thickness ($d=\xi_s$). $\bar\tau_{\rm sf}$ is the effective value for the spin-flip rate, given by Eq.~\eqref{bartau}.

The critical exchange field is suppressed by magnetic impurities as shown in Fig.~\ref{critical field magnetic impurities}(b). If the spin-flip rate is strong enough, the superconductor is driven to the normal state. For example, for an exchange field value of $\bar{h}=0.5T_{c0}$ and a spin-flip scattering of $(\bar\tau_{\rm sf}T_{c0})^{-1}=1$ the system would be in the normal state. This is reflected in the DoS shown in panel (a).

In the following, we focus on the magnetic impurity-free limit, $\tau_{sf}^{-1}=0$. In Fig.~\ref{multimode} we show the $h(T)$ diagrams for different values of the thicknesses of the superconducting layer. In order to study the effect of the thickness of the superconductor on the transition, we define a dimensionless exchange field as $ha/(\xi_s T_{c0})$, so that the normalizing factor is thickness-independent. As shown in the left panel of Fig.~\ref{multimode}: the thicker the sample, the higher the critical exchange field. The exchange field is located at the FI/S interface, so the influence of the interaction will diminish as we move away from the interface. Thus, the exchange field required to induce a phase transition increases monotonically with the thickness of the sample. Notably, there is a region where the critical temperature is a double-valued function of the field where, as explained above, the phase transition is of the first order for the lower temperatures. As the thickness of the sample is increased, the maximum critical exchange field is shifted towards lower temperatures, and the temperature range in which a first order transition occurs is reduced accordingly.

For thicknesses larger than a certain value $d^*$, the critical exchange field diverges at low temperatures, which means that the sample is in the superconducting state at $T=0$ for all values of the interface exchange field. In other words, if the sample is thick enough, the exchange field cannot induce a phase transition to the normal state. Therefore, depending on the thickness of the S layer, our numerical analysis suggests that the phase transition at zero temperature is either of the first order or does not take place.  

The right panel of Fig.~\ref{multimode} shows the critical thickness at which the phase transition occurs for different values of the exchange field. The metal is in the superconducting state if its thickness lies above the phase boundary.  In the absence of an exchange field the metal is in the superconducting state for temperatures lower than $T_{c0}$ and in the normal state for higher temperatures, regardless of the thickness.

The value of the critical thickness increases as the value of the exchange field is increased. As shown in Fig.~\ref{multimode}(b), the $h$ dependence of the critical thickness becomes weaker for large exchange fields, so that the curves approach a limiting behavior for large fields. By analyzing the $T \rightarrow 0$ limit, we obtained that the maximum thickness for which the critical exchange field exists and a phase transition occurs is (see Appendix~\ref{low temperature appendix} for the derivation)
\begin{equation}
    d^*=\sqrt{\frac{\gamma_E}{2}}\pi\xi_s \approx 3.0\xi_s\; ,
\end{equation}
where $\gamma_E \approx 1.781$ is the exponent of the Euler–Mascheroni constant. Note that the value $d^*$ of the critical thickness was already reported by Fominov \textit{et al.}~\cite{Fominov:2002} for an all metallic ferromagnet/superconductor bilayer at $T \rightarrow 0$ for a certain combination of junction parameters. We have demonstrated that this is also valid for a FI/S junction. If the thickness of the S layer is further increased, the critical exchange field no longer diverges at $T=0$, but at a higher temperature.

In the next section we present experimental results and contrast them with the previous theoretical analysis.

\section{Experiments and discussion}\label{Comparison to experiments}

In this section, we compare the experimental results of Al/EuS bilayers with our model's predictions.  To do that, we have fabricated four Al/AlO$_\mathrm{x}$/Al/EuS/silica tunnel junctions (from top to bottom). The distribution of the layers in the fabricated tunnel junctions is shown in Fig.~\ref{FI/S scheme}. A ferromagnetic insulator (EuS) layer is grown on the polished fused silica substrate. The bottom Al-wire (denoted as S in the picture) is in contact with the EuS layer, forming the FI/S interface. Two top Al-wires are rotated $90^\circ$ with respect to the bottom, these wires form the superconducting tunneling probes used to perform the spectroscopy measurements. The barrier between the two Al layers is made of nonstoichiometric aluminum oxide (AlO$_\mathrm{x}$). A thickness of the layers was monitored via quartz microbalance that was initially calibrated by means of the X-ray reflectivity measurements.

We fabricated four samples, with two junctions each, where we performed tunneling spectroscopy measurements at different temperatures with and without external magnetic fields in all samples. The tunneling spectroscopy of the junctions was done at temperatures down to 30 mK in a filtered cryogen-free dilution refrigerator. The I-V curves were measured in a standard dc four-wire configuration~\cite{Strambini:2017}, from which the differential conductance was calculated via numerical differentiation. From the value of the experimentally measured $T_c$ we can estimate the superconducting coherence length. Previous studies on thinner Al layers, 4 nm, showed larger values of the critical temperature  ($T_c\sim2.3$ K)\footnote{This unusual behavior of $T_c$ as a function of thickness is a known  property of Al thin films~\cite{Chubov:1968}}~\cite{Miao:2007}. In those cases the coherence length was estimated as $\xi_s \sim 35$ nm. Our films exhibit a smaller $T_c$ and hence we expect  a larger value of $\xi_s$. We can  then fairly assume that  in all our samples $d\ll \xi_s$ such that  fields  are  homogeneous in the Al adjacent to the EuS layer, as discussed in previous sections. The bottom Al-wire thickness was different for the four analyzed samples, see Table \ref{table samples}. The top Al wires were 10-12 nm thick, depending on the sample.
 
\begin{table}[H]
\begin{center}
 \begin{tabular}{| c | c | c | c | c | c |} 
 \hline
 Sample & $d$ (nm) &  $\Delta_{2,0}$ ($\mu$eV) & $\bar{h}$ ($\mu$eV) & $d_c$ (nm) \\ [0.5ex]
 \hline
  S1 & 10 & 245 & 78.5 & 4.5 \\ 
 \hline
 S2 & 8 & 255 & 118.5 & 5.3 \\ 
 \hline
 S3 & 9 & 245 & 146 & 7.6 \\ 
 \hline
 S4 & 3 & 0 & No SC & 5.4 \\ [1ex]
 \hline
 \end{tabular}
\caption{Properties of the Al layer in contact with the EuS for the different fabricated  samples. The averaged exchange field is extracted from the data at zero magnetic field. The last column gives the critical thickness obtained from Eq. (\ref{dcrit_CC}). See main text for details.\label{table samples}}
\end{center}
\end{table}

\begin{figure*}
    \centering
    \includegraphics[width=0.99\textwidth]{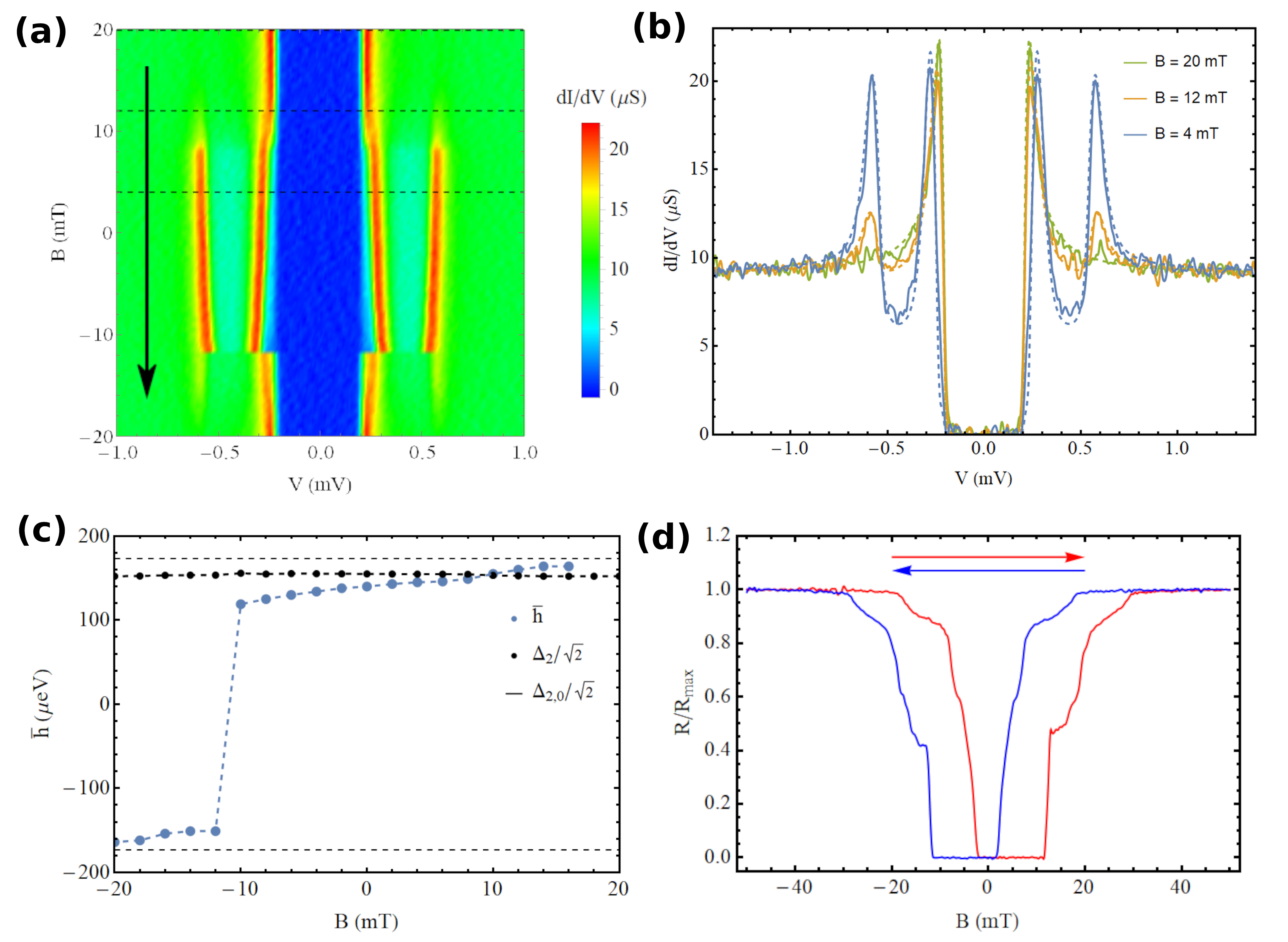}
        \caption{(a) The differential conductance of the S3 sample measured at 30 mK as a function of the external magnetic field and the voltage drop across the junction.  The arrow indicates the direction of the magnetic field sweep. At $B=10$ mT most of the bottom Al film undergoes a  transition into the superconducting state. 
        (b) $dI/dV$-curves at three different values of $B$ ($B=4$, $B=12$ and $B=20$ mT), indicated by dashed lines in (a).  
        The solid lines correspond to the experimental data, whereas the dashed lines to the theoretical fitting. The values of the fitting parameters used in the theoretical model at $B=0$ are $\bar h=146\; \mu$eV, $\Delta_2=219\; \mu$eV and $\tau_\mathrm{sf}^{-1}=38.5\; \mu$eV. 
        (c) Effective exchange field, $\bar h$,   and self-consistent order parameter, $\Delta_2$  of the bottom Al layer divided by $\sqrt{2}$. The horizontal dashed lines represent the Chandrasekhar-Clogston critical  field $\Delta_{2,0}/\sqrt{2}$ in the absence of magnetic impurities. 
        (d) Magnetoresistance of the bottom Al film adjacent to the EuS layer measured at $T=30$ mK.  }\label{meas15_S3}
\end{figure*}

Samples S1, S2, S3 show the characteristic large spin-splitting of the Al layer in contact with the EuS, even at zero applied magnetic field. The typical $dI/dV$ obtained from I-V characteristics are shown in Fig.~\ref{meas15_S3}(b) (we show only data for sample S3). Sample S4 does not show a spin-split $dI/dV$ because the bottom Al layer  is not superconducting. As we explain next, this is due to its small thickness.

By fitting the $dI/dV$-curves we can  extract the values of the two Al gaps, $\Delta_{1,2}$ (top and bottom Al layers, respectively), the effective exchange field, $\bar h$~\cite{meservey1994spin} (see Table \ref{table samples}) and the spin-flip rate $1/\bar \tau_{\rm sf}$.  The gap of the top Al layer, $\Delta_1$, can be determined from the $dI/dV$ at  large magnetic fields when $\Delta_2$ is suppressed [see green curve in Fig.~\ref{meas15_S3}(b)]. For example for S3 we obtain  $\Delta_{1,0} \approx 235\; \mu$eV. We determine from similar fitting the values of $\Delta_2$ and $\bar h$ at zero field: $\Delta_{2,0} \approx 245\; \mu$eV, $\bar h\approx 146\; \mu$eV. Here, $\Delta_{1,0}$ and $\Delta_{2,0}$ stand for the field-free order parameter. The real value of the order parameter is a bit smaller due to the exchange interaction. For example, at $B=0$, a self-consistent calculation gives a value of $\Delta_2=219\; \mu$eV, as shown in Fig.~\ref{meas15_S3}(c). In order to get a good agreement between the experimental and theoretical $dI/dV$, the effect of the magnetic impurities on the interface between EuS and the bottom Al layer is taken into account, with $\tau_\mathrm{sf}^{-1} \approx 38.5\; \mu$eV. This corresponds to a spin relaxation time $\tau_\mathrm{sf}\sim 10$ ps, much smaller than pure Al films. Provided that raw aluminum pellets used to grow the films are of 99.99 grade, the intrinsic level of impurities is below $0.01\%$. Another source of magnetic impurities could be a diffusion of Eu on the interface. A recent paper unveiled that epitaxial films of aluminum grow without intermixing on the single crystal EuS, demonstrating atomically sharp interfaces~\cite{Liu:2020}. Our samples feature polycristalline Al and EuS films, therefore it was difficult to demonstrate experimentally that the interfaces are equally sharp. Nevertheless, our X-ray photoelectron spectroscopy data (not shown) reveal rapid attenuation of the Eu peaks with increasing of thickness of the Al overlayer, thus corroborating negligible level of magnetic impurities in the Al and proving that the main source of spin flipping is the interface. Because of the tunneling barrier, the top Al is not affected by the exchange field and therefore it has the usual BCS DoS.

All experimental data  shown in Figs.~\ref{meas15_S3}(a-b) were  obtained in the low temperature regime, $T=30$ mK such that $T\ll\Delta_2$. In this regime, according to the discussion in Sec. \ref{critical field section}, one expects a first order phase transition when $\bar{h}\approx\Delta_0/\sqrt{2}$ when the spin-flip rate $1/\tau_{sf}$ is sufficiently small.  Using as a guide the $\bar h$ values obtained from the zero-magnetic field fitting and taking into  account the averaged thickness of the bottom Al layer we calculate the critical thickness below which superconductivity should  vanish from 
\begin{equation}
    d_c=\sqrt{2}d \bar{h}/\Delta_{2,0}\; .
    \label{dcrit_CC}
\end{equation}
The last column in Table \ref{table samples} shows  $d_c$ for all four samples. For the S1-S3 samples $d_c$ is close to, but smaller than the nominal thicknesses $d$ and the coexistence between superconductivity and spin-splitting is allowed. 

As we mentioned before, the sample S4 does not show the spin-splitted $dI/dV$ curve, and only the superconducting behaviour of the top Al layer is detected. Since the four samples were prepared one after the other, we can assume that the EuS layer for this sample is similar to the other samples, and then we can fairly assume a similar interfacial exchange and, therefore we can calculate a critical thickness of $d_c=5.4$ nm. This value exceeds the nominal Al thickness grown experimentally, and explains why superconducting transition may not take place. In other words, these results confirm that when the Al thickness is smaller than $d_c$, a superconducting transition does not take place as discussed in previous sections.

Further conclusions can be drawn from the dependence of the $dI/dV$-curves on an external magnetic field $B$.  Such field affects superconductivity in two ways: via Zeeman and  orbital effects. Because all films in our samples are very thin, and the magnetic field is applied in-plane we  can neglect the orbital effects~\cite{Fulde:1973} and focus on the dominant Zeeman interaction. The saturation magnetization of the EuS and parameters of the magnetization reversal are determined by the fabrication process of the layers [see the magnetization curve of Fig.~\ref{app_mag_inh_s1}(a)]. On the other hand, the interfacial exchange field that depends on the magnetic moment is controlled by external magnetic field. In Fig.~\ref{meas15_S3}(a-b) we show the full evolution of the measured $dI/dV$-curves  of sample S3 when the field is varied continuously from $+20$~mT to $-20$~mT. At high positive magnetic fields, the EuS is homogeneously magnetized, the exchange field is maximized, and   superconductivity cannot develop in the lower Al layer. Thus, the $dI/dV$ curve is  basically  proportional to the DoS of the upper Al layer, see green curve in Fig.~\ref{meas15_S3}(b), and the measured gap corresponds to the top Al layer ($\Delta_1$). From Fig.~\ref{meas15_S3}(a) one sees that at  $B\approx 10$~mT, a second coherence peak appears. This indicates that the Al layer in contact with the EuS goes through a phase transition into the superconducting state. 
When the field changes sign, it is opposed to magnetization. Parts of the EuS film start to switch their magnetization weakening the average exchange field.  This leads to the reduction of the  spin-splitting up to $B\approx-12$~mT. At this value we observe  a sudden disappearance of the outer coherence peak. This value of the $B$-field corresponds to the switching of the EuS magnetization. The magnetic film is now almost homogeneously magnetized in the opposite direction and the  resulting exchange field is strong enough to suppress superconductivity in the bottom layer.  The $dI/dV$ reflects  again the DoS of the Al at the top.

From our theoretical model we performed a fitting of several  $dI/dV$-curves at different magnetic fields, see for example Fig.~\ref{meas15_S3}(b). From these  fittings we extract  the values of $\bar h$ and $\Delta_2$, as shown in Fig.~\ref{meas15_S3}(c). In the absence of magnetic disorder one expects the first order phase transition at $\bar{h}_c=\Delta_{2,0}/\sqrt{2}$. 
For S3 this would correspond to  $\bar{h}_c \sim 173\; \mu$eV, shown in Fig.~\ref{meas15_S3}(c) with a dashed  horizontal line. 
However, as mentioned before, the magnetic disorder in our samples is sizable and leads to a smaller value of $\Delta_2$ even at zero splitting field. This explains their smaller values, shown with black dots in Fig.~\ref{meas15_S3}(c).
The values of $\bar h$ extracted from the fitting are shown with blue circles. $\bar h$ and $\Delta_2/\sqrt{2}$ cross at $B=10$ mT, the field at which the main phase transition  occurs (see Fig.~\ref{meas15_S3}(a)). This suggests that the critical exchange field is indeed equal to $\Delta_{2,0}^*/\sqrt{2}$, where now $\Delta_{2,0}^*$ is the value of the gap at zero exchange field but in the presence of magnetic impurities.

There is an additional important feature in the results of Fig.~\ref{meas15_S3}. On the one hand, the gap is practically constant for all the field values below the critical exchange field, indicating a first-order phase transition. On the other hand, Fig.~\ref{meas15_S3}(a) exhibits traces of the outer peak at values of $B$ larger than the critical one ($B>10$~mT). That peak disappears smoothly, as shown in Fig.~\ref{meas15_S3}(a-b). 
This indicates that for fields $10<B<16$~mT, the bottom Al layer is in a mixed phase, exhibiting  superconducting and normal  regions. This scenario is very likely given by small non-uniformity in the large cross-section of the junction,  200~$\mu$m$\times$200~$\mu$m. 
We quantify the mixed phase regime by a parameter $u$ which is equal to 1 in the complete  superconducting state and 0 in the normal state. The measured  differential conductance is then the result of the average:
${dI}/{dV}=u {dI}/{dV}|_{SS}+(1-u){dI}/{dV}|_{SN}$, where $SS$ ($SN$)  denotes the conductance measured when the bottom Al layer is in the superconducting (normal) state. 
Using the parameter $u$ we have fitted the $dI/dV$ curve at $B=12$ mT and $B=20$ mT, yellow and green curves in Fig.~\ref{meas15_S3}(b), by assuming $u=0.25$ and $u=0$ respectively. The agreement between theory and experiment is very  good.

Two complementary scenarios can  explain the mixed superconducting-normal {state}.  A possible explanation for the smooth disappearance of the coherence peaks is that the bottom Al has a spatially non-constant thickness. Fluctuations between 0.5-1 nm, will result in parts of the samples that turn superconducting at field values at which  other parts remain normal, {\it cf.} parameters for S3 in Table I. 
An additional possible scenario to explain the appearance of the  mixed phase  is the polycrystalline nature of the EuS films \cite{Strambini:2017,Miao:2009}. Indeed our EuS film  may contain at the interface with Al,   grains of   different sizes exhibiting  superparamagnetism. At low temperatures, their magnetization direction is basically determined by  the  interaction  with each other and the rest of the film via dipole and exchange interaction. As a consequence, one expects different values for the switching field, depending on  their  sizes and total  magnetic moment. 
The  distribution of grains with different magnetization directions causes a spatially  inhomogeneous exchange field, which leads to a phase transition occurring at different  values of the B-field in different parts of the sample. 
The last scenario is also consistent with the smooth switching of the resistance of  the Al layer as a function of the $B$-field, as   shown in Fig.~\ref{meas15_S3}(d). The resistance changes in a step-like form, suggesting the presence of a few types of grains that switch their magnetization at different $B$-field values (a more detailed description of the magnetization reversal of the EuS film is presented in  Appendix~\ref{magnetic inhomogeneity}).

It is also clear from our analysis that the strong $B$-dependence of the splitting field observed in our samples cannot be explained by the pure Zeeman field. A magnetic field of 20 mT corresponds to Zeeman energy of $g\mu_B B/2=1.16$ $\mu$eV, where $g \approx 2$ is the Land\'{e} g-factor. This energy is more than an order of magnitude smaller than the (measured) decrease in the spin-splitting that accompanies a decreasing of the field from 20 mT to 0 mT [see Fig.~\ref{meas15_S3}(c)]. 
Such a strong variation of the spin splitting is only due to the intrinsic properties of the EuS/Al interface. From a microscopic perspective, the interfacial  exchange field $h$, and hence the splitting is determined by 
the spatial distribution of magnetic moments, and the  value of the coupling between these moments  and the spin of conduction band electrons in Al~\cite{Zhang:2019,PtEuS}. 
{We have studied how the growth conditions affect  these parameters. Results of this investigation are consistent with earlier work~\cite{Smits:2004} and will be published elsewhere. In general, we found that high vacuum conditions are crucial to protect the EuS layer from oxidization and to avoid a paramagnetic phase at the EuS/Al interface. On the other hand, the sublimation temperature of the EuS powder defines the stoichiometry of the resulting EuS films. Non-optimal conditions were yielding samples with excess of Eu that strongly increases both the interfacial exchange field and the spin-flip relaxation rate, and above a certain critical concentration quenches the superconductivity in the adjacent Al-wire. }

\section{Conclusion}\label{conclusion}

We have presented an exhaustive study of the role of the superconductor thickness on the spectral properties and the critical exchange field of a FI/S bilayer. We found that the exchange field produces a well-defined spin splitting for thin and intermediate sized superconductors.   The splitting is, however, suppressed away from the FI/S interface in thick superconductors. Moreover, the spin-splitting at the FI/S interface is well approximated by the effective exchange field~\eqref{barh} even for thick samples. We have also studied the effect of the spin-flip relaxation due to magnetic impurities localized at the interface on the spectral properties of the system. The spin-flip relaxation suppresses the spin-splitting and lifts the superconducting gap, reducing the critical exchange field.
We have also studied the nature of the superconducting phase transition as a function of the exchange field and the S layer thickness. We found that thicker samples favor second-order phase transitions. In other words, we show that the thickness of the superconducting layer determines the order of the phase transition. We have also found a temperature-dependent thickness above which the critical surface field diverges, and no phase transition occurs, regardless of the value of the exchange field.
 
One of the important conclusions is that the coexistence between superconductivity and spin-splitting fields in thin layers is subject to a subtle balance between the thickness of the film and the interfacial exchange field determined by the quality of the FI/S interface and its magnetic properties.  Variations of the order of a few nm in the thickness of the  S-film may prevent the appearance of superconductivity. 
 
We have contrasted our theoretical findings with experiments on  EuS/Al samples fabricated by us. Our samples show a  well defined spin-split density of states. One of the samples with the thinnest Al film shows no superconductivity at 30 mK, in agreement with our theoretical prediction for thin films. In the other samples with a robust superconducting state, we found clear evidence of a first-order phase transition. We also found evidence of a coexistence phase regime where some part of the sample is superconducting, while the rest stays in the normal state. This mixed phase is attributed to the spatial fluctuation of the Al thickness, and a grainy texture of the magnetic EuS/Al interface.
 
Our results provide important insights for the fabrication of FI/S structures essential for applications where the control of a spin-splitting in a superconductor is crucial \cite{linder2015superconducting, eschrig2015spin, Giazotto:2015,Heikkila:2018,DeSimoni:2018, virtanen2018majorana,manna2020signature,vaitiekenas2020zero}.  
Our analysis shows that both the thickness of the S layer and the interfacial exchange field, given by the interface quality, are important quantities that have to  be carefully chosen in the fabrication process. 

{\it Note added-.} Almost simultaneously with our work  there was another related manuscript Ref.\cite{khindanov2020topological} in which the authors studied a thin superconductor/  ferromagnetic insulator bilayer,  and explored the engineering of topological superconductivity in an adjacent  semiconducting nanowire. To compute the density of states and order parameter they employed a very similar approach based on the Usadel equation, as the one in the present manuscript.

\section*{ Acknowledgements}
We thank Jagadeesh Moodera for useful discussions. F.S.B. acknowledges funding by the Spanish Ministerio de Ciencia, Innovación y Universidades (MICINN) (Project FIS2017-82804-P). T.T.H. acknowledges funding from the Academy of Finland (Project number 317118). A.H. acknowledges funding by the Department of Education of the Basque Government (Ikasiker grant). C.G.O. acknowledges funding of the PhD fellowship from MPC foundation. S.K. acknowledges for the fellowship of the ICTP Program for Training and Research in Italian laboratories, Trieste, Italy. C.R. acknowledges support from  Gobierno Vasco (Grant Nr.IT 1255-19).
F.G. acknowledges the
European Research Council under the EU’s Horizon 2020 Grant
Agreement No. 899315-TERASEC for partial financial support.
We acknowledge funding  from  EU’s Horizon 2020 research and innovation program under Grant Agreement No. 800923 (SUPERTED). F. S. B. thanks the Institute of Solid State Theory at University of Münster for its kind hospitality.

\begin{appendices}
\numberwithin{equation}{section}

\section{Multimode method ansatz}\label{Multimode method ansatz}

To solve the linearized Usadel equation~\eqref{Usadel_linear1} we use the multimode method~\cite{Fominov:2002}.

$\hat{f}$ is diagonal in spin space because $\boldsymbol{h}$ points in a single direction, so it can be written as
\begin{equation}
    \hat{f}=i f_0 + f_3 \sigma_3\; .
\end{equation}
Here $f_0$ and $f_3$ describe singlet and triplet pairs, respectively.

From Eq.~\eqref{Usadel_linear1}, we obtain that the Usadel equations for $f_0$ and $f_3$ are
\begin{eqnarray}
    \label{usadelf}
    \xi_s^2 \pi T_{c0}\partial_{xx}^2 f_0-|\omega_n|f_0+\Delta=0\; ,\\
    \xi_s^2 \pi T_{c0}\partial_{xx}^2 f_3-|\omega_n|f_3=0\; .
\end{eqnarray}
The boundary conditions are
\begin{eqnarray}
    \partial_x \left.f_0\right|_{x=0}=\mathrm{sgn}(w_n)\kappa_h \left.f_3+6\kappa_{sf} f_0\right|_{x=0}\; ,\\
    \partial_x \left.f_3\right|_{x=0}=-\mathrm{sgn}(w_n)\kappa_h \left.f_0+2\kappa_{sf} f_3\right|_{x=0}\; ,
\end{eqnarray}
\begin{eqnarray}
    \partial_x \left.f_0\right|_{x=d}=0\; ,\\
    \partial_x \left.f_3\right|_{x=d}=0\; .\label{usadelf_last}
\end{eqnarray}

Since $\Delta$ is spin independent, the singlet component is an even function of $\omega_n$, $f_0(-\omega_n)=f_0(\omega_n)$, and the triplet component is an odd function of $\omega_n$, $f_3(-\omega_n)=-f_3(\omega_n)$~\cite{bergeret2005odd,Bergeret:2006}. Therefore, the self-consistency equation \eqref{self-consistency0} can be rewritten in terms of the singlet component
\begin{equation}\label{self-consistency}
    \Delta\ln{\frac{T_{c0}}{T}}=2\pi T\sum_{\omega_n>0}\left(\frac{\Delta}{\omega_n}-f_0\right).
\end{equation}

In the multimode approach, one seeks as solution to Eqs. (\ref{usadelf}-\ref{self-consistency}) in the form~\cite{Fominov:2002}
\begin{equation}
\begin{split}
    f_0(x,\omega_n)=F_0(\omega_n)\cos{\left(\Omega_0\frac{x-d}{\xi_s}\right)}\\
    +\sum_{m=1}^{\infty}F_m(\omega_n)\frac{\cosh{\left(\Omega_m\frac{x-d}{\xi_s}\right)}}{\cosh{\left(\Omega_m\frac{d}{\xi_s}\right)}}
\end{split}
\end{equation}
\begin{equation}
    f_3(x,\omega_n)=F(\omega_n)\cosh{\left(\Omega_z\frac{x-d}{\xi_s}\right)}
\end{equation}
\begin{equation}\label{Delta_multimode}
    \Delta(x)=\delta_0\cos{\left(\Omega_0\frac{x-d}{\xi_s}\right)}+\sum_{m=1}^{\infty}\delta_m\frac{\cosh{\left(\Omega_m\frac{x-d}{\xi_s}\right)}}{\cosh{\left(\Omega_m\frac{d}{\xi_s}\right)}}\; ,
\end{equation}
where $\Omega_z^2=\omega_n/(\pi T_{c0})$. This ansatz automatically satisfies the boundary conditions at $x=d$.

The relation between the variables $\delta_m$ and $F_m$ is derived from \eqref{usadelf}. Solving $F$ from the B.C. equations at $x=0$, we obtain a set of linear homogeneous equations for the order parameter amplitudes $\delta_m$:
\begin{equation}\label{Knm}
    K_{nm}\delta_m=0\; ,
\end{equation}
where
\begin{subequations}\label{K}
\begin{multline}\label{Ka}
    K_{n0}=\biggl{(}\frac{\Omega_0}{\xi_s}\sin{\frac{\Omega_0 d}{\xi_s}}-6\kappa_{sf}\cos{\frac{\Omega_0 d}{\xi_s}}\\
    -\frac{\kappa_h^2\cos{\frac{\Omega_0 d}{\xi_s}}\cosh{\frac{\Omega_z d}{\xi_s}}}{\frac{\Omega_z}{\xi_s}\sinh{\frac{\Omega_z d}{\xi_s}}+2\kappa_{sf}\cosh{\frac{\Omega_z d}{\xi_s}}}\biggr{)}/(\omega_n+\Omega_0^2\pi T_{c0}),
\end{multline}
\begin{multline}\label{Kb}
    K_{nm}=\biggl{(}-\frac{\Omega_m}{\xi_s}\tanh{\frac{\Omega_m d}{\xi_s}}-6\kappa_{sf}\\
    -\frac{\kappa_h^2\cosh{\frac{\Omega_z d}{\xi_s}}}{\frac{\Omega_z}{\xi_s}\sinh{\frac{\Omega_z d}{\xi_s}}+2\kappa_{sf}\cosh{\frac{\Omega_z d}{\xi_s}}}\biggr{)}/(\omega_n-\Omega_m^2\pi T_{c0})\; ,\\
    \qquad m \geq 1\; .
\end{multline}
\end{subequations}
The mode frequencies $\Omega_m$ are determined by the self-consistency equation (\ref{self-consistency}).

Numerically, one can obtain $\kappa_h$ from Eq. \eqref{Knm} by considering the same amount of Matsubara frequencies and modes. The critical exchange field is then obtained by solving the equation
\begin{equation}\label{K equation}
    \det \hat{K}=0\; .
\end{equation}
The system of Eqs.~\eqref{K equation} might have multiple solutions for the critical temperature. For a given exchange field and thickness of the superconductor, the critical temperature is determined as the largest solution~\cite{Fominov:2002}.

\section{Derivation of the thickness \texorpdfstring{$d^*$}{}}\label{low temperature appendix}

\begin{figure*}
    \centering
    \includegraphics[width=0.75\textwidth]{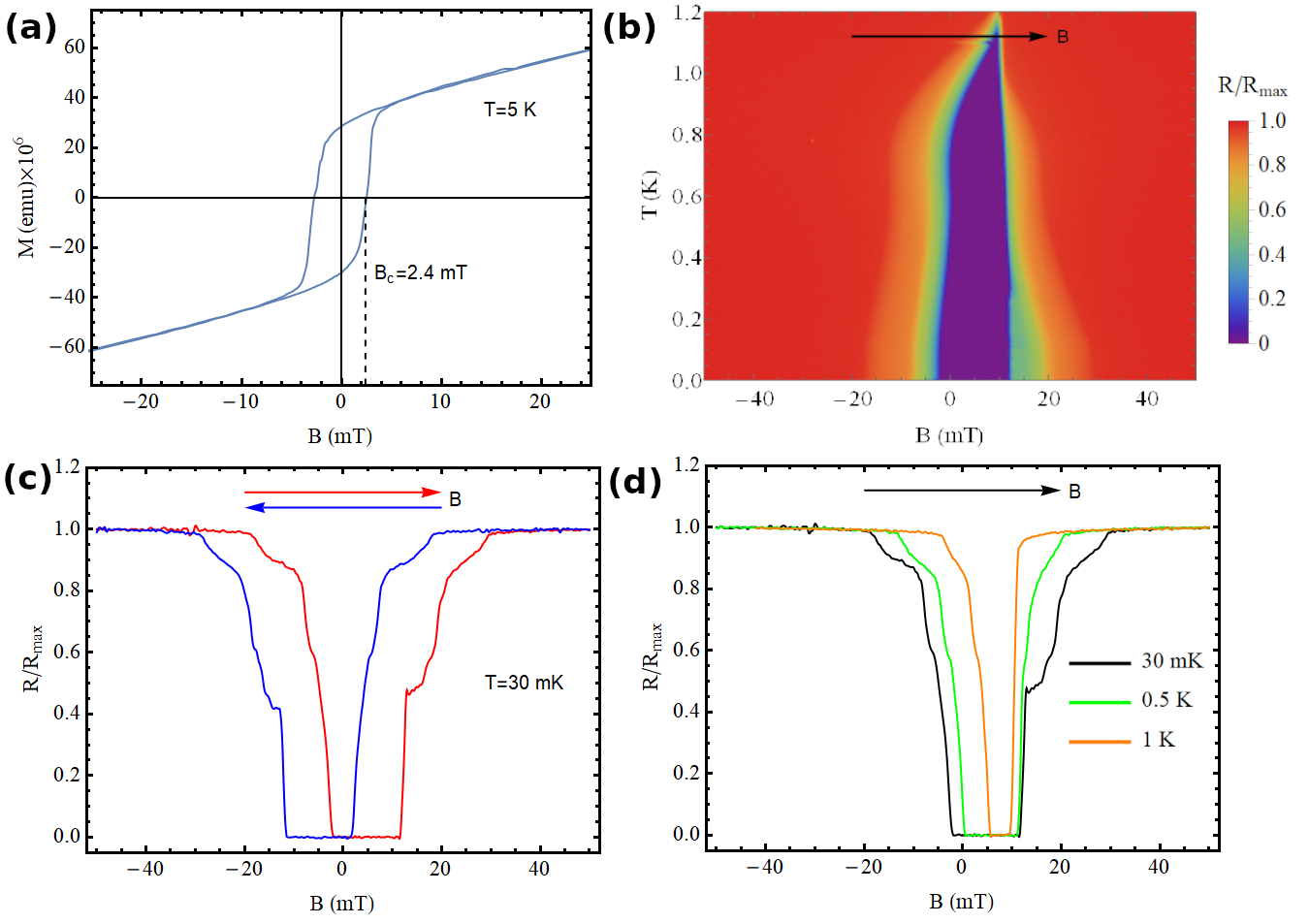}
        \caption{(a) Magnetization loop measured for a continuous EuS film at 5 K, with the field applied in the in-plane direction.  
        (b) Contour plot showing the resistance of Al wire adjacent to the EuS layer as a function of the external field and the temperature. The arrow indicates the sweep direction of the magnetic field. These measurements and the measurements shown in the next two panels were performed on the sample S3. (c) The hysteresis of the resistance of the Al wire adjacent to the EuS layer at $T=30$~mK. (d) The $B$-field dependence of the resistance of the Al wire adjacent to the EuS layer at different temperatures.  }\label{app_mag_inh_s1}
\end{figure*}

In this appendix we study the critical exchange field in the absence of magnetic impurities at $T=0$ and obtain an analytical expression for the thickness $d^*$ at which it diverges. The critical exchange field is obtained using the multimode method introduced in Sec.~\ref{critical field section}.

Taking the asymptotic expansion of the self-consistency equation~\eqref{self-consistency} in the $T \rightarrow 0$ limit, we obtain that the mode frequencies $\Omega_m$ are
\begin{gather}
    \Omega_0^2 \sim \frac{1}{2\gamma_{E}}\\
    \Omega_m^2 \sim \frac{T}{T_{c0}}(2m-1)\; , \quad m \geq 1\; ,
\end{gather}
where $\gamma_E \approx 1.781$ is the exponent of the Euler–Mascheroni constant.

Substituting this values in Eq.~\eqref{K}, we obtain that the elements of $\hat{K}$ to lowest order in $T/T_{c0}$ are
\begin{subequations}
\begin{equation}\label{Kn0 T=0_appendix}
    K_{n0} \propto 1-\frac{(2n+1)T}{T_{c0}}\left(1+\frac{\Omega_0d}{\kappa_h^2\xi_s^3}\tan{\frac{\Omega_0 d}{\xi_s}}\right)
\end{equation}
\begin{equation}
    K_{nm} \propto \frac{1}{(n-m+1)}\left(1+\frac{(2n+1)(2m-1)d^2}{\kappa_h^2\xi_s^4}\left(\frac{T}{T_{c0}}\right)^2\right)\; .
\end{equation}
\end{subequations}

The term $\tan{(\Omega_0 d/\xi_s)}$ in Eq.~\eqref{Kn0 T=0_appendix} diverges when $\Omega_0 d/\xi_s=\pi/2$. It can be proven that the value of $\Omega_0$ lies in the interval $0 < \Omega_0^2 < 1/(2\gamma_E)$, so the values of $d$ for which determinant~\eqref{K equation} diverges are
\begin{equation*}
    \sqrt{\frac{1}{2\gamma_E}}\frac{d}{\xi_s}\geq\frac{\pi}{2}\; .
\end{equation*}
Therefore, the maximum thickness for which the critical exchange field exists is
\begin{equation}
    d^*=\sqrt{\frac{\gamma_E}{2}}\pi\xi_s \approx 3.0\xi_s\; .
\end{equation}
For thicknesses greater than $d^*$, the critical exchange field will diverge at some finite temperature, so at $T=0$ the system will be in the superconducting state regardless of the magnitude of the exchange field.

\section{Magnetic properties  of the \texorpdfstring{E\lowercase{u}S}{} films}\label{magnetic inhomogeneity}

Magnetic properties of the thin films of EuS were studied in several works. To check the consistency of our data, we have grown a 12 nm thick film of EuS on the polished fused silica substrate under the same conditions used in the fabrication of the tunnel junctions. Figure~\ref{app_mag_inh_s1}(a) shows the magnetization loop measured at $5$~K in the magnetic field applied parallel to the film.    Magnetization reversal occurs at the coercive field, $2.4$~mT.  The film has a large value of the remanent magnetization similar to what was observed for the samples of Moodera's group~\cite{Miao:2009}. Nevertheless, magnetization grows slowly and saturates only in the field of a few Tesla. It implies that a polycrystalline EuS includes a certain amount of superparamagnetic grains weakly coupled to the main film. The relatively small value of the coercive field corroborates this observation. Indeed, for the magnetization reversal dominated by the domain walls movement, the coercive field increases with the size of the magnetic irregularities. Therefore, fine polycristalline structure results in  lower coercive fields~\cite{Miao:2009}. Comparing the magnetization loop presented in  Fig.~\ref{app_mag_inh_s1}(a) with the data reported in Ref.~\cite{Miao:2009}, it is clear that the  coercive field in  our sample is similar to the coercive field of  thin EuS films grown at 77~K which were found to have  grains with  4.4~nm diameters. Since EuS has no magnetocrystaline anisotropy, it is reasonable to expect very low values of the blocking temperature for the small superparamagnetic grains \cite{o1999modern}.
We could not measure magnetization loops below $1$~K. However, the dependence of the bottom Al layer's resistance interfaced with EuS film on the magnetic field provide indirect confirmation of this conjecture.   At $30$~mK, all superparamagnetic particles are in the blocked state. The $R(B)$-curve shows distinctive jumps that can be interpreted as magnetization reversal of the components with different magnetic anisotropy (Fig.~\ref{app_mag_inh_s1}(c)), as we explain next. 
If the exchange coupling between the grains is weak, the orientation of their magnetic moments depends on the dipolar interaction with the rest of the film and with the coupling with the external field. Whereas the external field will try to align the particles' moment, the dipolar interaction tries to orient the moment of the particles in the direction opposite to the magnetic moment of the film. Considering a descending branch of the resistance (red curve in Fig.~\ref{app_mag_inh_s1}(c)), we can see that all moments are aligned at a high positive field, and the effective exchange field is high (no superconductivity). When the field decreases down to zero, the dipolar interaction dominates. The resistance decreases in steps that correspond to consecutive switching  of the magnetization of the superparamagnetic particles. The average magnetic moment becomes smaller and leads to a decrease in the effective exchange field. This decrease allows the Al wire to become superconducting.  The zero-resistance state remains up to small negative values of the field. Further increase of the negative magnetic field leads again to consecutive reversal of the particles' magnetic moments accompanied by the switching of the total magnetic moment. The effective exchange field is large again and quenches superconductivity. This gives rise to the increase of the resistance. The same measurements performed at a higher temperature (Fig.~\ref{app_mag_inh_s1}(d)) show that the critical fields corresponding to the reversal of the superparamagnetic particles progressively disappear, showing the transition to the unblocked state.

\end{appendices}

\end{document}